\documentclass{webofc}
\usepackage[varg]{txfonts}
\usepackage{color}
\usepackage{amssymb}
\usepackage{amsmath}
\usepackage{array}
\usepackage{graphicx}
\usepackage{mathrsfs}
\newcommand{\MSbar}{\ensuremath{\overline{\text{MS}}}}
\begin{document}
\title{Deviation pattern approach for optimizing perturbative terms of QCD renormalization group invariant observables
}
%
%

\author{M.R.Khellat\thanks{mkhellat@gmail.com} and A.Mirjalili\thanks{a.mirjalili@yazd.ac.ir}
\\{\it{Physics Department, Yazd University, P.O.Box, 89195-741, Yazd, Iran}}}

\abstract{%
We first consider the idea of renormalization group-induced estimates, in the context of optimization procedures, for the Brodsky-Lepage-Mackenzie approach to generate higher-order contributions to QCD perturbative series. Secondly, we develop the deviation pattern approach (DPA) in which through a series of comparisons between lower-order RG-induced estimates and the corresponding analytical calculations, one could modify higher-order RG-induced estimates. Finally, using the normal estimation procedure and DPA, we get estimates of $\alpha_s^4$ corrections for the Bjorken sum rule of polarized deep-inelastic scattering and for the non-singlet contribution to the Adler function.
}
\maketitle
\section{Introduction}
\label{intro}
One of the main objectives of studying different ways of optimizing perturbative expansions for physical quantities is to disclose critical information about the sources of ambiguities that arise at different orders in perturbation theory studies. The limits and validity of these perturbative descriptions is another theoretical challenge which should be addressed  particularly in a complete optimization prescription. Consequently, unambiguous determination of the factorization scale would be crucial.  Light-front holographic formalism is an instance of a complete optimization prescription which defines an effective coupling for hadron dynamics at all momenta and takes advantage of the principle of maximum conformality (PMC)~\cite{pmc-1,Brodsky:2011ta,Brodsky:2013vpa} to fix the renormalization scale ambiguity. A review of this program alongside its current status and existing debates, such as the alternatives to PMC, can be found in~\cite{lfh-1}.

Moreover, important investigations have been going on for several years by the authors of~\cite{seblm-1, seblm-2, Kataev:2010du, scheme-dep-1, Kataev:2016aib} to understand the consequences of considering a conformal symmetry (CS) limit in QCD. The objective of these investigations involves: to find at first a plausible generalization of the Brodsky-Lepage-Mackenzie (BLM) approach which is capable of resumming charge renormalization contributions in QCD perrurbative series at different orders, and secondly clarify whether PMC, in the CS limit, generates series with scheme-independent coefficients. As a result of these investigations, an extension of the BLM approach, called sequential-BLM (seBLM), has been developed and evaluated in \cite{seblm-1, seblm-2}. The procedure of seBLM is formulated based on the proposed $\{\beta\}$-expansion for QCD observables and resums the $\beta$-dependent terms into a single renormalization scale through the renormalization-group equation (RGE) by rearranging these terms in several stages. It should be mentioned that the original version of PMC assumes a $\beta$-representation for the coefficients of QCD perturbative series which is different from the $\{\beta\}$-expansion of seBLM; however, PMC-II employs the same combination of $\beta$ coefficients as seBLM and performs a resummation of the $\beta$-pattern into different scales at different orders~\cite{pmc-2, pmc-3}.

In addition, to resolve the issue of scheme-invariance, PMC-II  suggests an analogy between the general structure of a QCD perturbative series and the induced terms in the structure of the series when the series is expressed in a ${\cal R}_{\delta}$ scheme~\cite{pmc-2}, as a subclass of the minimal subtraction schemes. The characteristic of this class of MS-like schemes is that they are related to each other through scale transformations. In fact, scheme-dependence of the BLM approach, as the predecessor of PMC, has been previously discussed in several cases~\cite{scheme-dep-2,scheme-dep-2n,scheme-dep-2nn}. Probably, PMC-II follows the strategy of identifying a class of plausible schemes, which translate their renormalization scales into each other, as a remedy to demonstrate a relation between scheme and scale ambiguities. However, this point should be considered cautiously even within the ${\cal R}_{\delta}$ schemes.

The next important theoretical criteria regarding any formulation of an optimization prescription is whether that formulation respects the general structure of perturbative expansions and the renormalization-group or not. A carefully established estimation procedure is potentially capable of revealing such characteristics of an optimization formulation. Motivated by the renormalization group equation, the idea of RG-inspired estimates was developed in~\cite{rg-est-1} to generate estimates for scheme-independent optimization prescriptions in QCD. We would devise a similar plan for the BLM scale-fixing procedure in which one would adopt a series of estimates for a renormalization group invariant observable at different orders up to ${\cal O}(\alpha_s^{n})$. After this stage, we would modify the $\sim \alpha_s^{i}, 2 \leq i \leq n$ estimate to produce an estimate at $\sim \alpha_s^{n+1}$. To generate an $(i+1)$-th order estimate, we start from the $i$-th order BLM scale-fixed series and evolve the series using the $(i+1)$-th order RGE.

\section{The procedure of estimation}
\label{sec-1}
Consider a RG invariant perturbative expansion in an MS-like scheme as $R(Q^2)=\;r_0+r_1 a_s(Q^2)+\sum_{i=2}^\infty r_i {a_s}^i(Q^2)$ for which the normalization $a_s={\alpha_s}/{\pi}$ has been chosen. At any order, this series is supposed to represent a measurable quantity within that order of perturbative approximation. On the other hand, due to the renormalization procedure, the truncated series has become a function of the renormalized parameters. To resolve the issue, one could take advantage of the BLM proposal to assign a single renormalization scale or multiple scales at different orders for the series. Here, we follow the single scale extension of BLM proposed in~\cite{blm-ext-1} and extend the corresponding definition of the BLM scale to any order through the following recursive relation~\cite{MKI}
\begin{equation}\label{trns1-3f}
\ell^{(k)} \equiv \ln \left( \mu_{BLM}^2/\mu_0^2 \right) = \ell^{(k-1)} + \sum_{i=0}^{k-2} c_{ki} a_p^{k-2} ( \mu_{BLM})\;.
\end{equation}
In Eq.{~(\ref{trns1-3f})} the NLO and N$^2$LO BLM scales and coefficients, corresponding to $k=2$ and $k=3$, can be adopted from~\cite{MKI}. Taking the N$^{(k-1)}$LO BLM scale-fixed series
\begin{equation*}\label{trns1}
R^{(k)}_{BLM} = r_0 + r_{1} a_p ( \mu_{BLM}) + \bar{r}_{2} a^2_p ( \mu_{BLM})+ \ldots + \bar{r}_{k} a^k_p ( \mu_{BLM}),
\end{equation*}
one could evolve the series using the RG equation at order $(k+1)$
\begin{align}\label{trns0}
\mu^2 \frac{{\rm d} a}{{\rm d}\mu^2} &= - \sum_{i=0}^{k-1} \beta_i a(\mu)^{i+2},\nonumber \\
a (\mu_0) &= \exp \left(-\ell^{(k)} \beta(a) \partial_{a} \right) a \mid_{a=a(\mu_{BLM})}~,
\end{align}
where the operator representation of \cite{seblm-1} has been adopted to formulate the evolution of  the renormalized coupling through the renormalization scales.

Here, we would formulate our procedures based on the $n_f$ expansion $r_i=\sum_{k=0}^{i-1} r_{ik}n_f^k$ for the coefficients of the RG-invariant expansion. However, we should note that any implementation of BLM should obey its core principle to resum vacuum polarization contributions, which are accumulated in the $\{\beta\}$-coefficients of the renormalization scheme. The trick to take care of vacuum polarization insertions by resumming flavor-dependent terms would not work at N$^4$LO and beyond where $n_f^1$ contributions start to appear which are not related to charge renormalization. As a result, our proposed formulation is restricted to be valid up to N$^4$LO.

 N$^2$LO and N$^3$LO estimates equivalent to $k=3$ and $k=4$ can be found in Eqs.(1,8)  from ~\cite{MKI}. For instance, the analytical \emph{SU}$(N_c)$ expression for the N$^2$LO estimate would be
\begin{align}\label{r3pre}
r_{3}^{(est)}= &- \frac{121}{16}\frac{C_{A}^2}{T_F^2}\frac
{r_{21}^2}{r_1}\;-\;\frac{17}{8}\frac{C_{A}^2}{T_F}r_{21}
\;-\; \frac{11}{2}\frac{C_{A}}{T_F}\frac{r_{20} r_{21}}{r_1}\nonumber\\
&+\; \left( 2 \frac{r_{20} r_{21}}{r_1} + \frac{5}{4} r_{21} C_{A} +
\frac{3}{4} r_{21} C_{F} \right) n_f + \frac{r_{21}^2} {r_{1}}
n_f^2\;.
\end{align}
Here $[ T^{a} , T^{a}]_{ij}=C_F \delta_{ij}$ and $f^{acd}f^{bcd}=C_A \delta^{ab}$ are quadratic Casimir operators of the fundamental and adjoint representations of the color group \emph{SU}$(N_c)$ and tr$(T^a T^b)=T_F \delta^{ab}$ is the trace normalization of the fundamental representation; $\{C_F=\frac{N_c^2-1}{2N_c}, C_A=N_c, T_F=1/2\}$. As it is also explained in the next section, $\sim a^3$ and $\sim a^4$ estimates in ~\cite{MKI} are generated by the inverse of Eq.~(\ref{trns0}) as an operator equation at ${\cal O}(\alpha_s^{3})$ and ${\cal O}(\alpha_s^{4})$-levels incorporating $\ell^{(1)}$ and $\ell^{(2)}$, respectively. The third-order estimation is given by Eq.~(\ref{r3pre}) in which the presence of $-\ell^{(1)}= -3 r_{21}/ T_F r_1$ is evident. It should also be noted that, in this framework, transition from $n_f$ expansion to seBLM $\{\beta\}$-pattern would not be possible.

The most important characteristic of these N$^2$LO and N$^3$LO estimates is that they vanish completely at the perturbative quenched QED (pqQED) limit $\{C_F=1, C_A=0, T_F=1, n_f=0\}$. We refer the reader to a detailed study of the pqQED model, the specifications of the conformally invariant limit of the model. Its connection with the consideration of massless perturbation theory results, obtained in pqQED can be found in~\cite{Kataev:2013vua}.

\subsection{Deviation pattern approach}
\label{sec-2-1}
For generating the exact N$^{k}$LO results from a N$^{k-1}$LO BLM scale-fixed series, we need $\ell^{(k)}$ and $\bar{r}_k$; however, we just have access to $\ell^{(k-1)}$ and the estimates are generated on the basis of $\bar{r}_k=0$. In other words, estimates are naturally deviated from the exact results. On the other hand, these deviations occur at all orders and, for purely numerical purposes, we can take advantage of our knowledge of the deviation of lower-order estimates from the exact results. A possible algorithm which could directly include deviations in the estimation procedure by performing numerical comparisons and modifications would be,
\begin{enumerate}
\item $k=2$ estimation and modification:
\begin{enumerate}
\item generate the N$^2$LO estimate via Eq.~(\ref{r3pre})
\item make the comparison $r_3^{(est)}=r_3^{(exact)}$ which is equivalent to the following three equations
\begin{equation*}
\begin{aligned}
&r_{30}^{(exact)}\;=\;r_{30}^{(est)}\;=\;- \frac{121}{16}\frac{C_{A}^2}{T_F^2}\frac{r_{21}^2}{r_1}\;-\;\frac{17}{8}\frac{C_{A}^2}{T_F}r_{21}
\;-\; \frac{11}{2}\frac{C_{A}}{T_F}\frac{r_{20} r_{21}}{r_1}~,\nonumber\\
&r_{31}^{(exact)}\;=\;r_{31}^{(est)}\;=\;2 \frac{r_{20} r_{21}}{r_1} + \frac{5}{4} r_{21} C_{A} + \frac{3}{4} r_{21} C_{F},\;\; r_{32}^{(exact)}\;=\;r_{32}^{(est)}\;=\; \frac{r_{21}^2} {r_{1}}~.
\end{aligned}
\end{equation*}
\item solve the three equations for $\{ r_1, r_{20}, r_{21}\}$ and substitute the solutions in the expression for $r_4^{(est)}$; the modification is like changing the weights of the constituents of $r_4^{(est)}$.
\end{enumerate}
\item $k>2$ estimations and modifications (suppose we have access to exact results up to N$^k$LO ):
\begin{enumerate}
\item generate N$^2$LO to N$^k$LO estimates, i.e. $k-1$ estimates.
\item make the comparisons $\{ r_j^{(est)}=r_j, 3 \leq j \leq k+1\}$ which are equivalent to $3+4+ \ldots+k=(1/2) (k-1)(k+4)$ equations $\{ r_{mn}^{(est)}=r_{mn}^{(exact)}\}$ for $3 \leq m \leq k+1$ and $0 \leq n < m$.
\item perform a selection of equations because the number of free parameters controlling the estimates is $(k/2)(k+1)$; therefore, $(k-2)$ comparisons should be left out (for an explanation of the point in 4th order estimations, see section 2.2 in~\cite{MKI}).
\item adapt $(k/2)(k+1)$ parameters obtained from the previous step to modify $r_{k+2}^{(est)}$.
\end{enumerate}
\end{enumerate}

\section{Adler function and Bjorken sum rule}
\label{sec-3}
The primary building block of the Adler function and $R$-ratio is the vacuum polarization function $\Pi (Q^2)$ which is a scale-dependent object in QFT and in the \MSbar-scheme is a function of the renormalization scale $\mu$ and the running coupling $a_s$
\begin{align*}
(q_{\mu}q_{\nu} - q^2 g_{\mu \nu}) \Pi(Q^2) = i \int
{\mathrm{d}}^4 x e^{iq.x} \langle 0 \mid T[j_{\mu}(x) j_{\nu}(0)]
\mid 0 \rangle~.
\end{align*}
Here $Q^2=-q^2$ and the time-dependent correlator is responsible for the production of the vacuum polarization. The Adler function and $R$-ratio are related to the vacuum polarization function and the hadronic EM vacuum polarization function as follow~\cite{adler-74, Baikov:2008jh}
\begin{align}\label{Adler.prodI}
&D(Q^2) =-12 \pi^2 Q^2 \frac{\mathrm{d}}{\mathrm{d}Q^2} \Pi(Q^2)~, \nonumber \\
&\tilde{R}(s) = 6 \pi \left( \Pi^{EM} (-s-i\epsilon)) - \Pi^{EM}(-s+i\epsilon) \right)~.
\end{align}
These two functions are related to each other through the well-known integral transformations
\begin{equation}\label{adler.trans.}
D(Q^2)=\int_0^{\infty} \frac{Q^2  \tilde{R}(s) \; \mathrm{d}s}{(s + Q^2)^2}, \;\; \tilde{R}(s)=\frac{i}{2\pi}\;\int_{s - i\epsilon}^{s + i\epsilon} \frac{\mathrm{d}z}{z} D_{pt}(-z)\;.
\end{equation}
The corresponding perturbative expansions $D_{pt}(Q^2)= \sum_n d_n^{pt} \alpha_s^n(Q^2)$ and $\tilde{R}(s)=\sum_n \tilde{r}_n \alpha_s^n(s)$ would also become connected by introducing appropriate analytical continuation procedures~\cite{shirkov-00}.

The Bjorken polarized sum rule is an integration over the difference between proton and neutron polarized structure functions\cite{Pasechnik:2008th}
\begin{align*}
\Gamma_1^{p-n} &= \int_{0}^{1} \mathrm{d}x \; \left[g_1^p(x,Q^2) - g_1^n(x,Q^2)\right] \nonumber \\
&=\frac{g_A}{6} C^{Bjp}(Q^2) + \sum_{j=2}^{\infty}
\frac{\mu_{2j}^{p-n}(Q^2)}{Q^{2j-2}}~.
\end{align*}
$g_A$ represents the charge of the axial vector current of the nucleon. The non-singlet coefficient function $C_{ns}^{Bjp}$ and the non-singlet Adler function $D^{ns}$ would have the following perturbation theory expansions in the \MSbar-scheme,
\begin{align}
&D^{ns}(a_s) = 1 + d_1 a_s + d_2 a_s^2 + d_3 a_s^3 + d_4 a_s^4 + {\cal O}(a_s^5)~,\\
&C_{ns}^{Bjp} (a_s) = 1 + c_1 a_s + c_2 a_s^2 + c_3 a_s^3 + c_4 a_s^4 + {\cal O} (a_s^5)~.
\end{align}
It would be convenient to compare the exact results for the color structures at N$^2$LO approximation with the color structures of the respective estimated contributions to these functions, i.e. $d_3$ and $c_3$ with $d_3^{est}$ and $c_3^{est}$. We could write the corresponding differences as follows:
\begin{align}
\label{d3-d3est}
d_3 - d_3^{est} = &\underline{\underline{-\frac{69}{128} C_F^3}} + C_F^2 T_F n_f \left[ \frac{15}{64} + \frac{17}{4} \zeta_3 - \underline{\underline{5\zeta_5}} \right] +C_F T_F^2 n_f^2 \left[\frac{119}{432} + \frac{14}{9}\zeta_3 - \underline{\frac{4}{3}\zeta_3^2} \right]\\ \nonumber
&+ C_F^2 C_A \left[ -\frac{133}{128} - \frac{77}{8} \zeta_3 + \underline{\underline{\frac{55}{4}\zeta_5}} \right] + C_F C_A T_F n_f \left[ -\frac{924}{432} - \frac{329}{36}\zeta_3 + \underline{\frac{22}{3}\zeta_3^2} + \underline{\underline{\frac{5}{6}\zeta_5}} \right]\\ \nonumber
&+ C_F C_A^2 \left[ \frac{24569}{6912} + \frac{407}{36}\zeta_3 - \underline{\frac{121}{12}\zeta_3^2} - \underline{\underline{\frac{55}{24}\zeta_5} }\right]~, \\
\label{c3-c3est}
c_3-c_3^{est} = & \underline{\underline{-\frac{3}{128} C_F^3}} + C_F^2 T_F n_f \left[ \frac{155}{576} - \underline{\underline{\frac{5}{12}\zeta_3}} \right] + C_F T_F^2 n_f^2 \left[ -\frac{43}{216} \right] + C_F^2 C_A \left[ -\frac{145}{576} - \underline{\underline{\frac{11}{12}\zeta_3}} \right]\\ \nonumber
&+ C_F C_A T_F n_f \left[ \frac{1339}{864} + \underline{\underline{\frac{3}{4}\zeta_3 - \frac{5}{6}\zeta_5}} \right] + C_F C_A^2 \left[ -\frac{2143}{864} + \underline{\underline{\frac{55}{24}\zeta_5}} \right]~.
\end{align}
The terms which are double-underlined do exclusively belong to the exact analytical results $d_3$ and $c_3$ while the terms proportional to $\zeta_3^2$ in eq.~(\ref{d3-d3est}) belong to $d_3^{est}$. Transcendental Riemann functions related to the Bjorken sum rule of the polarized lepton-hadron DIS $C^{Bjp}_{ns}$ start to appear at $\sim a^3$ while for $D^{ns}$ they exist at $\sim a^2$; this is the reason why there are no $\zeta_3$ and $\zeta_5$ contributions in $c_3^{est}$.

Having an input on the difference between N$^2$LO estimates of $D^{ns}$ and $C^{Bjp}_{ns}$ and the corresponding exact results, it would be possible to follow the estimation procedure of Sec.~\ref{sec-1} and the deviation pattern approach in Sec.~\ref{sec-2-1} to produce N$^3$LO estimates for these functions in Table~\ref{tab-1}. The $est$ superscript refers to the method of Sec.~\ref{sec-1} and $est1$ superscript to the DPA estimates. As listed in the table, one can also find the exact numerical results for both $D^{ns}(Q^2)$~\cite{Baikov:2008jh} and $C^{Bjp}_{ns}(Q^2)$~\cite{Baikov:2010je} at the fourth-order of perturbative approximation.
\begin{table}
\centering
\caption{$C^{Bjp}_{ns}$ and $D^{ns}$ $\sim a^4$ estimates}
\label{tab-1}
\begin{tabular}{l l l l l l l l l l l} 
\hline
$C^{Bjp}$ & $n_f^0$ & $n_f^1$ & $n_f^2$ & $n_f^3$ & $D^{NS}$ & $n_f^0$ & $n_f^1$ & $n_f^2$ & $n_f^3$ \\ [0.5ex] 
\hline \\
$c^{est}_{4}$ & -265.4 & 95.26 & -5.94 & 0.08    & $d^{est}_{4}$ & 362.1 & -99.08 & 5.04 & -0.05 \\ 
$c^{est1}_{4}$ & -261.3 & 67.71 & -3.62 & 0.04  & $d^{est1}_{4}$ & 317.2 & -93.26 & 4.76 & -0.03
\\
$c^{exact}_{4}$ & -479.4 & 123.4 & -7.69 & 0.10 & $d^{exact}_{4}$ & 407.4 & -103.3 & 5.63 & -0.03 \\ [1ex] 
\end{tabular} \\
\end{table}
\newline

\emph{Acknowledgment} \;\; M.K thanks Andrei Kataev for the very useful discussions and his comments on several topics related to the work done here. He is also grateful for the financial support provided by JINR and the organization committee of the XXIII International Baldin Seminar on High Energy Physics Problems which was dedicated to the 90th anniversary of Academician A. M. Baldin. The authors acknowledge the comments of the respected referee which improved and clarified parts of the paper.
%
%
%

\end{document}